# Ferromagnetic Superconductivity in Two-dimensional Niobium Diselenide

Tingyu Qu[1,2], Shangjian Jin[2,3], Fuchen Hou[4], Deyi Fu[3], Junye Huang[5], Darryl Foo Chuan Wei[3], Xiao Chang[5], Kenji Watanabe[6], Takashi Taniguchi[6], Junhao Lin[4], Shaffique Adam[2,3,5,7], Barbaros Özyilmaz[1,2,3,5,8,*]

[1] NUS Graduate School, Integrative Sciences and Engineering Programme (ISEP), National University of Singapore, Singapore, Singapore.

[2] Department of Physics, National University of Singapore, Singapore, Singapore.

[3] Centre for Advanced 2D Materials, National University of Singapore, Singapore, Singapore.

[4] Department of Physics, Southern University of Science and Technology, Shenzhen, China.

[5] Department of Materials Science and Engineering, National University of Singapore, Singapore, Singapore.

[6] National Institute for Materials Science, 1-1 Namiki, Tsukuba 305-0044, Japan.

[7] Yale-NUS College, 16 College Ave West, Singapore, Singapore.

[8] Institute for Functional Intelligent Materials (I-FIM), National University of Singapore, Singapore, Singapore.

[*] e-mail: barbaros@nus.edu.sg

The co-existence of ferromagnetism and superconductivity becomes possible through unconventional pairing in the superconducting state[1]. Such materials are exceedingly rare in solid-state systems but are promising platforms to explore topological phases[2],[3], such as Majorana bound states[4]–[6]. Theoretical investigations date back to the late 1950s[7],[8], but only a few systems have so far been experimentally identified as potential hosts[9]–[13]. Here, we show that atomically-thin niobium diselenide ($NbSe_2$) intercalated with dilute cobalt atoms spontaneously displays ferromagnetism below the superconducting transition temperature ($T_C$). We elucidate the origin of this phase by constructing a magnetic tunnel junction that consists of cobalt and cobalt-doped niobium diselenide (Co-$NbSe_2$) as the two ferromagnetic electrodes, with an ultra-thin boron nitride as the tunnelling barrier. At a temperature well below $T_C$, the tunnelling magnetoresistance shows a bistable state, suggesting a ferromagnetic order in Co-$NbSe_2$. We propose a Ruderman–Kittel–Kasuya–Yosida exchange coupling mechanism based on the spin-triplet superconducting order parameter to mediate such ferromagnetism. We further perform non-local lateral spin valve measurements to confirm the origin of the ferromagnetism. The observation of Hanle precession signals show spin diffusion length up to micrometres below $T_C$, demonstrating an intrinsic spin-triplet nature in superconducting $NbSe_2$. Our discovery of superconductivity-mediated ferromagnetism opens the door to an alternative design of ferromagnetic superconductors.

## Introduction

Ferromagnetic superconductors display intrinsic co-existence of ferromagnetism and superconductivity[1]. Though the nature of Cooper pairs in such materials is still under debate[14]–[16], a spin-triplet pairing state is generally accepted as a necessary ingredient[1],[11]. So far, only a few systems have been shown to meet this requirement. For example, in uranium-based compounds and stacked graphene, Pauli limit violation is used as the hallmark of spin-triplet Cooper pairs[11]–[13]. However, an intriguing question remains as whether a superconducting condensate could spontaneously display ferromagnetism.

Recently, transition metal dichalcogenide (TMD) superconductors have emerged as an alternative candidate to realize such unconventional superconductivity. Few-layer TMDs naturally host anisotropic pairing channels, which support additional symmetry breaking and unusual superconducting properties such as Ising protection[17]–[19], spin-triplet pairing states[18]–[21] and topological superconductivity[22]–[25]. In the family of TMD superconductors, niobium diselenide ($NbSe_2$) stands out because of its high superconducting transition temperature ($T_C$)[17] and the expected sizeable singlet-triplet mixing[19],[20]. In addition, as predicted, when intercalated with magnetic atoms, a spin-triplet superconducting host could give rise to Ruderman–Kittel–Kasuya–Yosida (RKKY)-mediated ferromagnetism[26]. In this paper, we use spin-dependent transport measurements to show that below $T_C$, $NbSe_2$ becomes ferromagnetic when intercalated with dilute cobalt (Co) atoms.

We develop two distinct device configurations. First, we construct a vertical magnetic tunnel junction (MTJ) with evaporated Co and Co-doped $NbSe_2$ (Co-$NbSe_2$) as the two ferromagnetic electrodes and BN as the tunnelling layer in-between. We study superconductivity-mediated ferromagnetism via spin-dependent tunnelling measurements inside the gap of $NbSe_2$. Second, we utilize the ferromagnetic Co-$NbSe_2$ electrodes as a lateral spin injector/detector for pristine $NbSe_2$ channels and study magnetic switching and Hanle precession via non-local spin valve (NLSV) measurements. These results directly confirm that $NbSe_2$ hosts spin-triplet pairing states and that it is the former that gives rise to ferromagnetism in Co-$NbSe_2$ below $T_C$. Our system is promising for studying the co-existence of ferromagnetism and superconductivity and building emerging quantum devices in two-dimensional (2D) limit.

## Device Design and Magnetic Intercalation

We select ultra-thin (2 ~ 3 atomic layers) hexagonal boron nitride (BN) to encapsulate $NbSe_2$. Patterned Co lines are deposited onto the BN/$NbSe_2$ heterostructure via e-beam evaporation and form the ferromagnetic contacts. Here, the ultra-thin BN plays two additional roles. At device level, it acts as a spin dependent tunnelling barrier. More crucially, through our fabrication process, it also allows for interstitial doping of $NbSe_2$ (See Methods) (Fig. 1a, left panel). As evident from the mass contrast scanning transmission electron microscopy (STEM) image, no obvious Co clusters are observed (Fig. 1b, top panel). Based on the electron energy loss spectroscopy (EELS), the Co doping concentration ($n_{\mathrm{Co}}$) decays exponentially with an average value of around 3% and a penetration depth of around a-few nm (Fig. 1b, bottom panel). We conclude that few-layer BN likely serves as a stable buffer layer assisting the intercalation of Co of van der Waals (VdW) structure (Extended Data 1).

The magnetic properties are characterized by three different transport measurements, namely, a four-probe charge transport measurement to characterize the Co-doped $NbSe_2$ channel resistance, a two-probe magnetic transport measurement of the tunnelling magnetoresistance (TMR) in a magnetic tunnel junction (MTJ) consisting of Co/BN/Co-$NbSe_2$, and a non-local lateral spin valve measurement with Co-$NbSe_2$ as an injector/detector and undoped $NbSe_2$ as the channel (Fig. 1a, right panel and Extended Data 2). We start our discussion with the four-probe measurement to understand the impact of Co intercalation on the superconductivity of multi-layer $NbSe_2$. Encouragingly, the superconducting state is preserved suggesting that the Co doping in $NbSe_2$ is sufficiently dilute (See Supplementary Information). Further, voltage–current ($V-I$) measurements in the vicinity of superconducting transition (around $T_{\mathrm{C}}$) follows a power-law ($V{\sim}I^{\alpha}$) and its non-linearity becomes more prominent with decreasing temperature (Fig. 1c). This can be well fitted by the theoretical model of Berezinskii–Kosterlitz–Thouless (BKT) transition[27],[28], implying that our Co-doped superconducting layers are in the 2D regime. Such an observation is surprising because the thicknesses of our samples are all around 10 ~ 12 nm (Extended Data 3). It suggests that the Co intercalation decouples the bulk $NbSe_2$ without breaking the superconductivity. In fact, the decoupling of the superconducting layers becomes more prominent with increasing Co concentration. For such samples, despite a reduction of the zero-field $T_{\mathrm{C}}$, the in-plane upper critical field ($B_{\mathrm{C2||}}$) is enhanced and can exceed the Pauli limit ($B_{\mathrm{Pauli}}$) (Fig. 1d and Supplementary Information). The $B_{\mathrm{C2||}}/B_{\mathrm{Pauli}}$ vs. $T_{\mathrm{C}}$ relation of our Co-$NbSe_2$ behaves similarly to that of few-layer $NbSe_2$ flakes[17], which also agrees with the recent discovery of reduced dimensionality in an intercalated bulk $NbSe_2$ achieved by a different means[29]. We also studied devices where the BN buffer layer was transparent or missing altogether. In neither devices, we observed a superconducting transition.

**Superconductivity-mediated Ferromagnetism**

Next, we study the phase diagram of our Co-$NbSe_2$ (Device A) with temperature and magnetic field as two independent variables. In pristine $NbSe_2$, the field-free superconductivity persists at all temperatures below $T_\mathrm{C}$. In contrast, at $B = 0$, although our Co-$NbSe_2$ exhibits a superconducting state below $T_\mathrm{C0}$ (~ 6.6 K), a resistive phase re-emerges below a second critical temperature $T_\mathrm{K}$ ($T_\mathrm{K} < T_\mathrm{C0}$) such that at base temperature (1.5 K), the resistance increases to nearly 40% of that just above $T_\mathrm{C}$. Surprisingly, this unusual phase can be completely screened out under a finite magnetic field ($B_\mathrm{K}$), where our system re-enters the non-resistive state (re-entrant superconductivity) (Fig. 2a). We also notice that $T_\mathrm{K}$ is anisotropic under the in-plane and out-of-plane field directions (Extended Data 4). The resistance anomaly triggered by the superconducting transition resembles a failed superconductor and hints to a fermionic phase induced in a bosonic host[30],[31]. However, our system differs in one critical aspect, namely, the existence of a non-resistive state at intermediate temperature and magnetic field ranges ($T_\mathrm{K} < T < T_\mathrm{C}$ and $B_\mathrm{K} < B < B_{\mathrm{C2}||}$) for a finite range of $n_\mathrm{Co}$ (See Extended Data 5 and Supplementary Information). We will later argue that $n_\mathrm{Co}$ corresponds to a length scale comparable to the superconducting coherence length for ferromagnetism to emerge inside the gap.

Having confirmed that the resistance anomaly results from the Co intercalation, we investigate the magnetic properties of this state. Here, Co and Co-$NbSe_2$ form the two ferromagnets in our MTJs and we study their voltage-bias dependent tunnelling spectrum at fixed temperatures and fixed fields. At low temperature and small field, we observe a prominent zero-bias peak (ZBP) in the differential conductance ($dI/dV$). Notably, both the temperature range and the field range are consistent with the resistive state in the phase diagram occurring at $T < T_\mathrm{K}$ and $B < B_\mathrm{K}$. At temperatures above $T_\mathrm{K}$ and magnetic fields above $B_\mathrm{K}$, the ZBP vanishes, which likewise matches the conditions for the non-resistive phase in Fig. 2a. To further understand the origin of ZBP, we fit the tunnelling spectrum using the $p$-wave pairing model[32]. It captures well both the magnitude of the ZBP and the observed critical temperature at $T_\mathrm{K}$. These results hint towards mediated magnetism inside the superconducting gap[33] (See Methods and Supplementary Information). To rule out any spurious origins that could arise from the Co contact[34],[35], we turn our attention to tunnelling magnetoresistance (TMR) measurements of the Co/BN/Co-$NbSe_2$ MTJ (Fig. 3a).

We first measure TMR as a function of $B_{||}$ along the easy axis of the Co electrode. We ensure a constant current ($\leq 0.1$ μA) through the junction such that the potential in Co-$NbSe_2$ falls inside gap. Below $T_K$, we observe hysteretic switching of the TMR with a magnitude ($\Delta R$) that gradually increases with decreasing temperature such that it saturates at 500 Ω ($\Delta R/R \sim$ 10%) at base temperature ($T = 1.5$ K) (See top panel in Fig. 3b, Fig. 3c, and top panel in Fig. 3d). This demonstrates the existence of two ferromagnets of distinct coercive fields[36]. Using anisotropic magnetoresistance (AMR) measurements on the Co contact alone, we identify that the higher switching field originates from the Co electrode (Extended Data 6). Therefore, the second switch at lower field must result from the switching of Co-$NbSe_2$ (Fig. 3b, bottom panel). Critically, at temperatures above $T_K$ but still below $T_C$, no bistable states in the TMR are observed (Extended Data 7). Note that the observation of an anti-parallel configuration also rules out proximity-induced ferromagnetism in $NbSe_2$ as the proximity effect would spontaneously favour a parallel configuration. Next, we fix the temperature below $T_K$ and study the bias-dependence of the TMR in more detail (Fig. 3e). The bias current determines the potential difference across the tunnel junction (Fig. 3d, bottom panel), thus it can be used to tune the potential relative to the superconducting gap. Only for bias currents that confine the potential deep inside the gap, a clear hysteretic switching in TMR is observed.

We now present a phenomenological model based on our observation that the ferromagnetism only occurs below $T_K$ and when the bias is inside the superconducting gap. Since the ferromagnetic order only exists with dilute Co doping, we can rule out direct exchange coupling. We also note that the effect only appears above a certain threshold of Co doping, namely, a minimum separation distance between the Co atoms is needed. This suggests indirect exchange coupling, namely, RKKY interaction. Above $T_C$, RKKY should be suppressed for two reasons in the metallic state. First, the Fermi velocity ($v_F$) is high, leading to an oscillation of the RKKY coupling strength ($J_{RKKY}$)[37],[38]. Second, the strong intrinsic spin orbit coupling leads to rapid dephasing[39]. Also, if the Fermi level is outside the superconducting gap, RKKY is dominated by filled electrons and the subsequent oscillating $J_{RKKY}$ suppresses the long-range ferromagnetic coupling. But below $T_C$ and inside the gap, the electrons are screened out and the RKKY can now be mediated by superconducting carriers. Recently, Parkin S. and co-workers have indeed reported the coupling of two individual magnetic moments (within 5-nm range) in superconducting niobium by scanning tunnelling microscopy measurement[40]. Theoretically, ferromagnetic RKKY in superconducting condensate prefers a spin-triplet ground state (Fig. 3f)[26]. In addition, for RKKY to emerge below $T_C$, the magnetic moment separation should be comparable to the superconducting coherence length ($\xi_{GL}$)[41]. Indeed, this condition is met when both temperature and magnetic field are below critical thresholds, such that $T < T_K < T_C$ and $B < B_K < B_{C2||}$, and with the bias deep inside the gap. Only then, $\xi_{GL}$ becomes smaller than the magnetic coupling length and the RKKY interaction can be mediated by sufficient superconducting carriers (Supplementary Information).

**Spin-triplet Pairing State**

Our model critically depends on the existence of a triplet state. While $NbSe_2$ has been reported to host such state, a direct proof is missing. Proximity-induced spin supercurrents[42],[43] cannot probe intrinsic triplet pairs in the superconductor. Our approach is to use Co-$NbSe_2$ instead as a superconducting ferromagnet to enable direct spin injection into the gap of $NbSe_2$. To do so, we build a lateral non-local spin valve with a few-layer $NbSe_2$ as the channel material (Device B) with three different channel lengths (Fig. 4a).

We first note that devices with either a direct Co contact or devices with a much thicker BN barrier (≥ 5 layers), limiting the Co intercalation, did not show any non-local signal (See Supplementary information). Only for device where $n_{Co}$ is around 3%, we observe spin signals in the superconducting state. Fig. 4b shows a 2D mapping of the non-local resistance ($R_{NL}$) as a function of $I_{Inj}$ and $B_{||}$ along the easy axis of Co-$NbSe_2$ (set at $-40$ mT and swept from 0 to $+40$ mT), where an anti-parallel can be clearly resolved where the largest switching magnitude of $R_{NL}$ ($\Delta R_{NL}$) is more than 150% of the background (Extended Data 8). Individual field sweeps are shown in Fig. 4c, where we plot $R_{NL}$ by scanning $B_{||}$ with different injection currents ($I_{Inj}$). We observe nearly symmetric hysteretic switching between two distinct resistance states at several representative $I_{Inj}$ provided the injection bias is inside the gap.

To unambiguously probe the spin-triplet state in $NbSe_2$, channel length-dependent Hanle measurements are performed, where the injector and detector are polarized by $B_{||}$ before an out-of-plane field ($B_{\perp}$) is swept for in-plane spin precession. First, we see nearly symmetric Hanle precession curves under $I_{Inj} = 0.1$ μA and $T = 100$ mK (Extended Data 9a). Second, the precession occurs only for temperatures below $T_C$ and when $I_{Inj}$ is inside the gap (Extended

Data 9b). Third, we observe a clear length dependence of the precession signals. Fig. 4d shows $R_{\mathrm{NL}}$ vs. $B_{\perp}$ as a function of channel lengths ($L_{\mathrm{CH}}$) equal to 0.6 μm, 0.8 μm and 3.0 μm. For $L_{\mathrm{CH}} = 0.8$ μm, the signal changes sign at $B_{\perp} = \pm 0.125$ T, and then saturates at $\pm 0.5$ T, showing a clear signature of spin precession. For $L_{\mathrm{CH}} = 3.0$ μm, the Hanle curve shows even two additional peaks at $\pm 0.4$ T, implying a 180° phase rotation of the spins. A second spin-phase rotation at longer channel length suggests the spin information is well protected in the superconducting channel. Using Bloch equations in the diffusive regime, we estimate the spin diffusion length ($\lambda_{\mathrm{S}}$) in the order of micrometres (See Methods and Supplementary Information). Such long-range spin transport through the superconducting channel rules out any proximity effect, and strongly suggest an intrinsic triplet-pairing state in $NbSe_2$.

## Summary and Outlook

In summary, we discovered superconductivity-mediated ferromagnetism via magnetic Co intercalation in atomically-thin $NbSe_2$. The ferromagnetism in the superconducting condensate occurs due to the presence of a spin-triplet pairing state, which is demonstrated by our spin-dependent transport measurements. Our work not only provides crucial complementary evidence of the spin-triplet nature in $NbSe_2$ and but also implies that the system is potentially suitable for device level implementation towards new quantum circuits, for example, as a 2D platform for harbouring Majorana bound states.

## References

[1] Aoki, D., Ishida, K., & Flouquet, J. Review of U-based ferromagnetic superconductors: Comparison between UGe2, URhGe, and UCoGe. *J. Phys. Soc. Japan* **88**, 022001 (2019).

[2] Sato, M., & Ando, Y. Topological superconductors: a review. *Rep. Prog. Phys.* **80**, 076501 (2017).

[3] Jiao, L., *et al*. Chiral superconductivity in heavy-fermion metal UTe2. *Nature* **579**, 523–527 (2020).

[4] N. Read, D. Green. Paired states of fermions in two dimensions with breaking of parity and time-reversal symmetries and the fractional quantum Hall effect. *Phys. Rev. B* **61**, 10267–10297 (2000).

[5] A. Y. Kitaev. Unpaired Majorana fermions in quantum wires. *Phys. Uspekhi* **44** (10S), 131–136 (2001).

[6] Alicea, J. New directions in the pursuit of Majorana fermions in solid state systems. *Rep. Prog. Phys.* **75**, 076501 (2012).

[7] Ginzburg, V. L. Ferromagnetic superconductors. *Zh. Eksp. Teor. Fiz.* **31**, 202 (1956).

[8] Anderson, P. W., & Suhl, H. Spin alignment in the superconducting state. *Phys. Rev.* **116**, 898 (1959).

[9] Saxena, S. S., *et al*. Superconductivity on the border of itinerant-electron ferromagnetism in UGe2. *Nature* **406**, 587–592 (2000).

[10] Aoki, D., *et al*. Coexistence of superconductivity and ferromagnetism in URhGe. *Nature* **413**, 613–616 (2001).

[11] Ran, S., *et al*. Nearly ferromagnetic spin-triplet superconductivity. *Science* **365**, 684–687 (2019).

[12] Cao, Y., Park, J. M., Watanabe, K., Taniguchi, T. & Jarillo-Herrero, P. Pauli-limit violation and re-entrant superconductivity in moiré graphene. *Nature* **595**, 526–531 (2021).

[13] Zhou, H., *et al*. Isospin magnetism and spin-polarized superconductivity in Bernal bilayer graphene. *Science* **375**, 774–778 (2022).

[14] Ishida, K., *et al*. Spin-triplet superconductivity in Sr2RuO4 identified by 17O Knight shift. *Nature* **396**, 658–660 (1998).

[15] Flouquet, J., & Buzdin, A. Ferromagnetic superconductors. *Phys. World.* **15**, 41 (2002).

[16] Agterberg, D. F. The symmetry of superconducting Sr2RuO4. *Nat. Phys.* **17**, 169–170 (2021).

[17] Xi, X., *et al*. Ising pairing in superconducting NbSe2 atomic layers. *Nat. Phys.* **12**, 139–143 (2016).

[18] Fatemi, V., *et al*. Electrically tunable low-density superconductivity in a monolayer topological insulator. *Science* **362**, 926–929 (2018).

[19] Wickramaratne, D., Khmelevskyi, S., Agterberg, D. F. & Mazin, I. Ising superconductivity and magnetism in NbSe2. *Phys. Rev. X* **10**, 041003 (2020).

[20] Hamill, A. *et al*. Two-fold symmetric superconductivity in few-layer NbSe2. *Nat. Phys.* **17**, 949–954 (2021).

[21] Lu, J., *et al*. Evidence for two-dimensional Ising superconductivity in gated MoS2. *Science* **350**, 1353–1357 (2015).

[22] Sohn, E., *et al*. An unusual continuous paramagnetic-limited superconducting phase transition in 2D NbSe2. *Nat. Mater.* **17**, 504–508 (2018).

[23] Fatemi, V., *et al*. Electrically tunable low-density superconductivity in a monolayer topological insulator. *Science* **362**, 926–929 (2018).

[24] Kezilebieke, S., *et al*. Topological superconductivity in a van der Waals heterostructure. *Nature* **588**, 424–428 (2020).

[25] Cho, C. W., *et al*. Nodal and nematic superconducting phases in NbSe2 monolayers from competing superconducting channels. *Phys. Rev. Lett.* **129**, 087002 (2022).

[26] Schmid, H., Steiner, J. F., Franke, K. J., & von Oppen, F. Quantum Yu-Shiba-Rusinov dimers. *Phys. Rev. B* **105**, 235406 (2022).

[27] Berezinskii, V. Destruction of long-range order in one-dimensional and two-dimensional systems possessing a continuous symmetry group. ii. quantum systems. *Sov. Phys. JETP* **34**, 610–616 (1972).

[28] Kosterlitz, J. M. & Thouless, D. J. Ordering, metastability and phase transitions in two-dimensional systems. *J. Phys. C: Solid State Phys.* **6**, 1181 (1973).

[29] Zhang, H., *et al.* Tailored Ising superconductivity in intercalated bulk NbSe2. *Nat. Phys.* **18**, 1425–1430 (2022).

[30] Kapitulnik, A., Kivelson, S. A., & Spivak, B. Colloquium: Anomalous metals: Failed superconductors. *Rev. Mod. Phys.* **91**, 011002 (2019).

[31] Yang, C., *et al.* Signatures of a strange metal in a bosonic system. *Nature* **601**, 205–210 (2022).

[32] Oh, M., et al. Evidence for unconventional superconductivity in twisted bilayer graphene. *Nature* **600**, 240–245 (2021).

[33] Sticlet, D., & Morari, C. Topological superconductivity from magnetic impurities on monolayer NbSe2. *Phys. Rev. B* **100**, 075420 (2019).

[34] Kezilebieke, S., Žitko, R., Dvorak, M., Ojanen, T., & Liljeroth, P. Observation of coexistence of Yu-Shiba-Rusinov states and spin-flip excitations. *Nano Lett.* **19**, 4614–4619 (2019).

[35] Žutić, I., & Valls, O. T. Tunneling spectroscopy for ferromagnet/superconductor junctions. *Phys. Rev. B* **61**, 1555 (2000).

[36] Julliere, M. Tunneling between ferromagnetic films. *Phys. Lett. A* **54**, 225–226 (1975).

[37] Kiss, T., *et al.* Charge-order-maximized momentum-dependent superconductivity. *Nat. Phys.* **3**, 720–725 (2007).

[38] Khestanova, E., *et al.* Unusual suppression of the superconducting energy gap and critical temperature in atomically thin NbSe2. *Nano Lett.* **18**, 2623–2629 (2018).

[39] Zyuzin, A. A., & Loss, D. RKKY interaction on surfaces of topological insulators with superconducting proximity effect. *Phys. Rev. B* **90**, 125443 (2014).

[40] Küster, F., Brinker, S., Lounis, S., Parkin, S. S., & Sessi, P. Long range and highly tunable interaction between local spins coupled to a superconducting condensate. *Nat. Commun.* **12**, 6722 (2021).

[41] Ghanbari, A. & Linder, J. RKKY interaction in a spin-split superconductor. *Phys. Rev. B* **104**, 094527 (2021).

[42] Robinson, J., Witt, J. & Blamire, M. Controlled injection of spin-triplet supercurrents into a strong ferromagnet. *Science* **329**, 59–61 (2010).

[43] Sanchez-Manzano, D., *et al.* Extremely long-range, high-temperature Josephson coupling across a half-metallic ferromagnet. *Nat. Mater.* **21**, 188–194 (2022).

[44] Dynes, R. C., Narayanamurti, V., & Garno, J. P. Direct measurement of quasiparticle-lifetime broadening in a strong-coupled superconductor. *Phys. Rev. Lett.* **41**, 1509 (1978).

[45] Tinkham, M. Introduction to superconductivity. Courier Corporation (2004).

[46] De Trey, P., Gygax, S., & Jan, J. P. Anisotropy of the Ginzburg-Landau parameter κ in NbSe2. *J. Low Temp. Phys.* **11**, 421−434 (1973).

[47] Dao, V. H., & Chibotaru, L. F. Destruction of global coherence in long superconducting nanocylinders. *Phys. Rev. B* **79**, 134524 (2009).

[48] Jedema, F. J., Heersche, H. B., Filip, A. T., Baselmans, J. J. A., & Van Wees, B. J. Electrical detection of spin precession in a metallic mesoscopic spin valve. *Nature* **416**, 713–716 (2002).

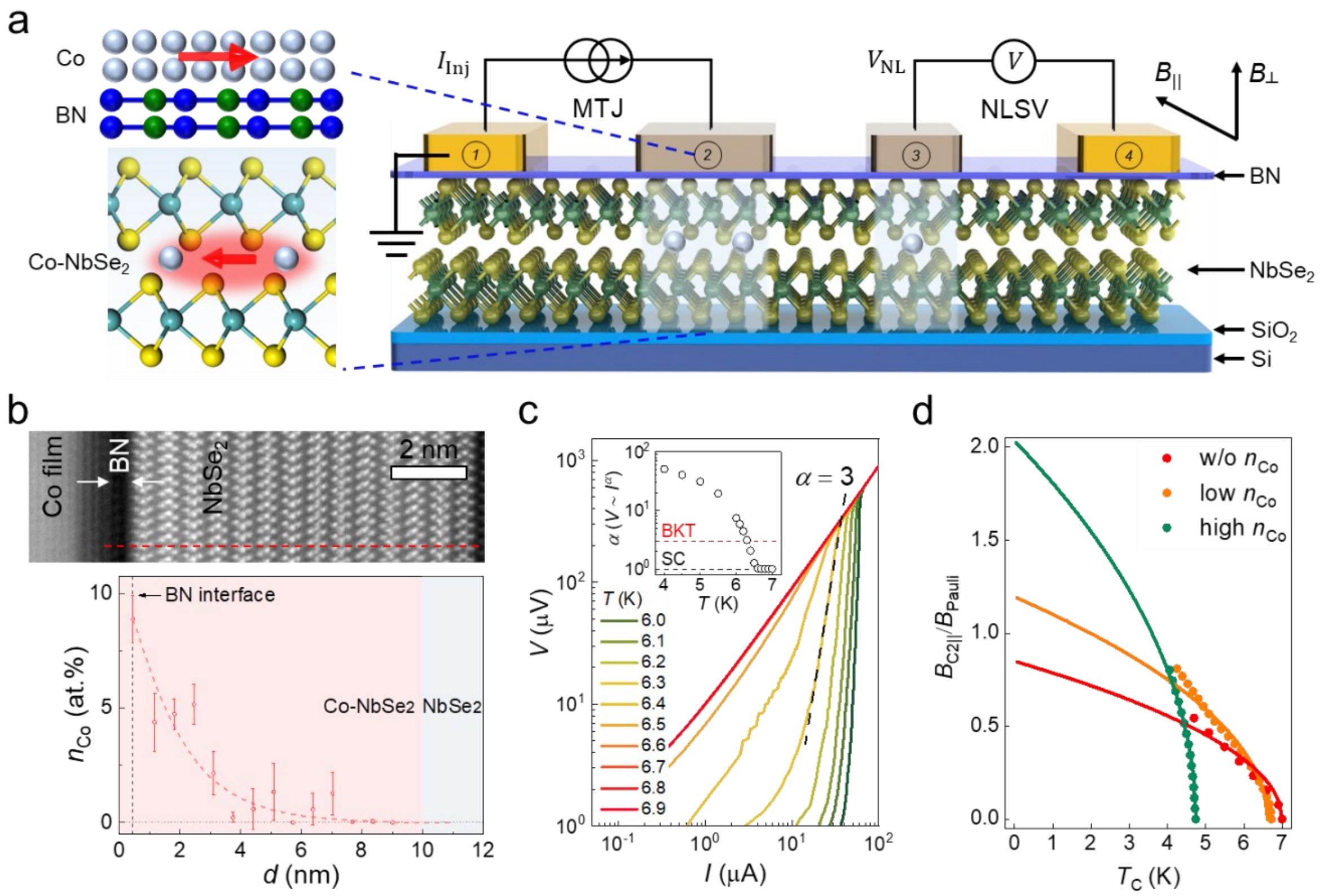

**Fig. 1 | Magnetic intercalation and 2D superconductivity. a**, Magnetic doping strategy and device measurement configuration. The ultra-thin BN acts as a diffusion barrier that allows Co intercalation in Co-$NbSe_2$. Magnetic contacts (Co, as labelled by #2 and #3) are used for spin injection and detection and non-magnetic contacts (Au, as labelled by #1 and #4) are used for ground and reference. The spin transport can be probed by both the local voltage (across contacts #1 and #2) and the non-local voltage (across contacts #3 and #4). The charge transport can be characterized by the channel voltage (across contacts #2 and #3) in a four-probe measurement. **b**, Scanning transmission electron microscopy (STEM) image of the Co/BN/Co-$NbSe_2$ heterojunctions (top panel) and the corresponding depth profiles of $n_{\mathrm{Co}}$ on the $NbSe_2$ side (bottom panel). **c**, $V-I$ curves for the Co-intercalated $NbSe_2$ devices with temperatures ranging from below $T_{\mathrm{C}}$ to just above $T_{\mathrm{C}}$. Inset: Temperature dependence of the exponent $\alpha$ deduced from the power-law, $V\sim I^{\alpha}$. As indicated by the black dashed line, $\alpha$ approaches 3 at $T = 6.3$ K. **d**, $B_{\mathrm{C2}||}/B_{\mathrm{Pauli}}$ vs. $T_{\mathrm{C}}$ for Co-intercalated $NbSe_2$ devices with different doping levels. The devices with low $n_{\mathrm{Co}}$, high $n_{\mathrm{Co}}$ and without (w/o) $n_{\mathrm{Co}}$ correspond to Devices A, C and E, respectively.

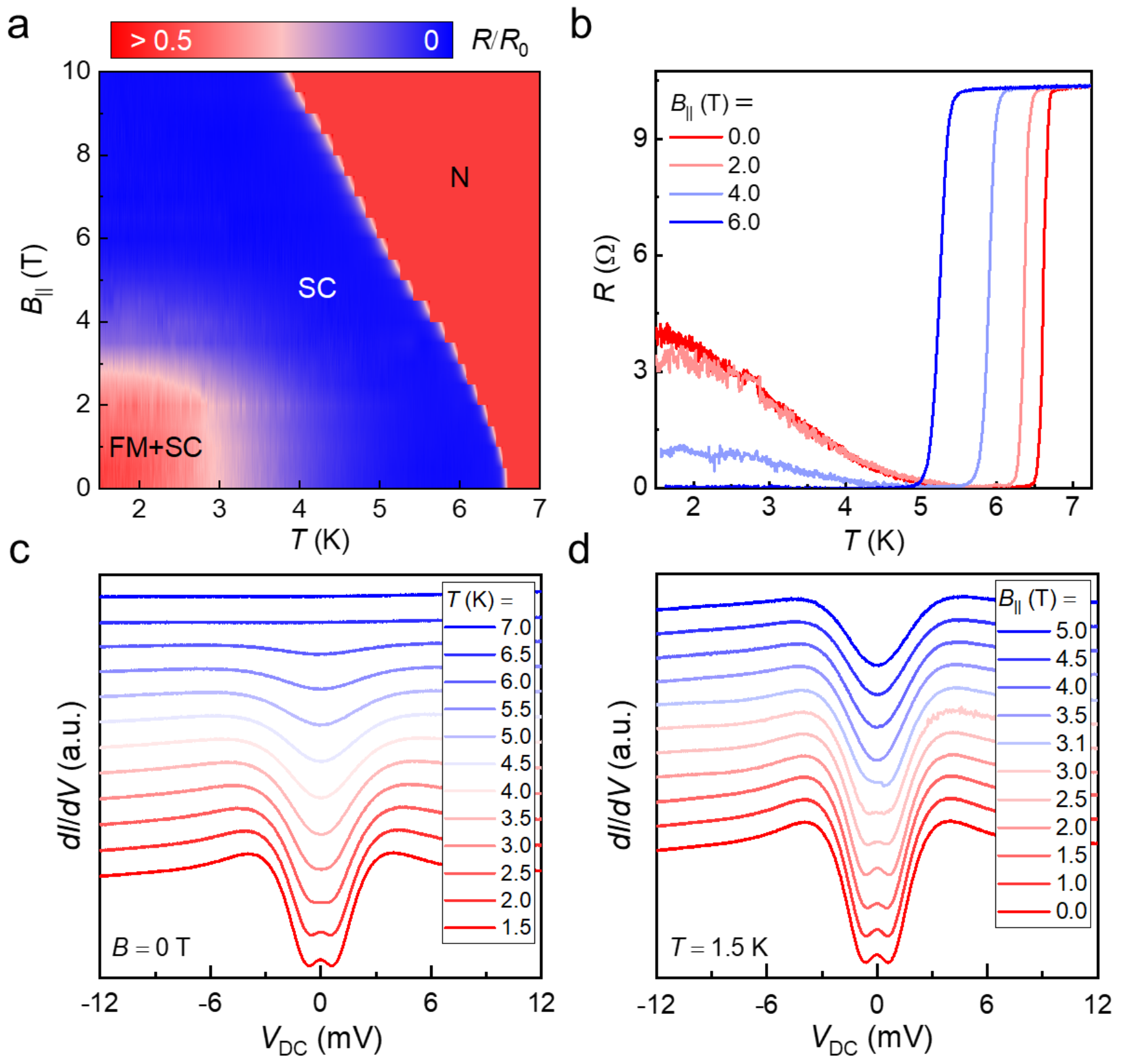

**Fig. 2 | Ferromagnetic phase below $T_C$ and the YSR states. a**, The $(B_{||}, T)$ phase diagram. Our system undergoes from a normal state (N) to a superconducting state (SC) at $T < T_C$. Under the condition of $T < T_K$ and $B < B_K$, a resistance anomaly phase occurs, corresponding to a ferromagnetic phase coexisting with superconductivity (FM+SC). **b**, $R - T$ curves by line cuts with discrete $B_{||}$ in (**a**). A constant AC current (1 μA) is applied through the channel. **c-d**, Two-probe $dI/dV$ curves with temperature (**c**) and field (**d**) sequences. The ZBP occurs under the condition $T < T_K$ and $B < B_K$.

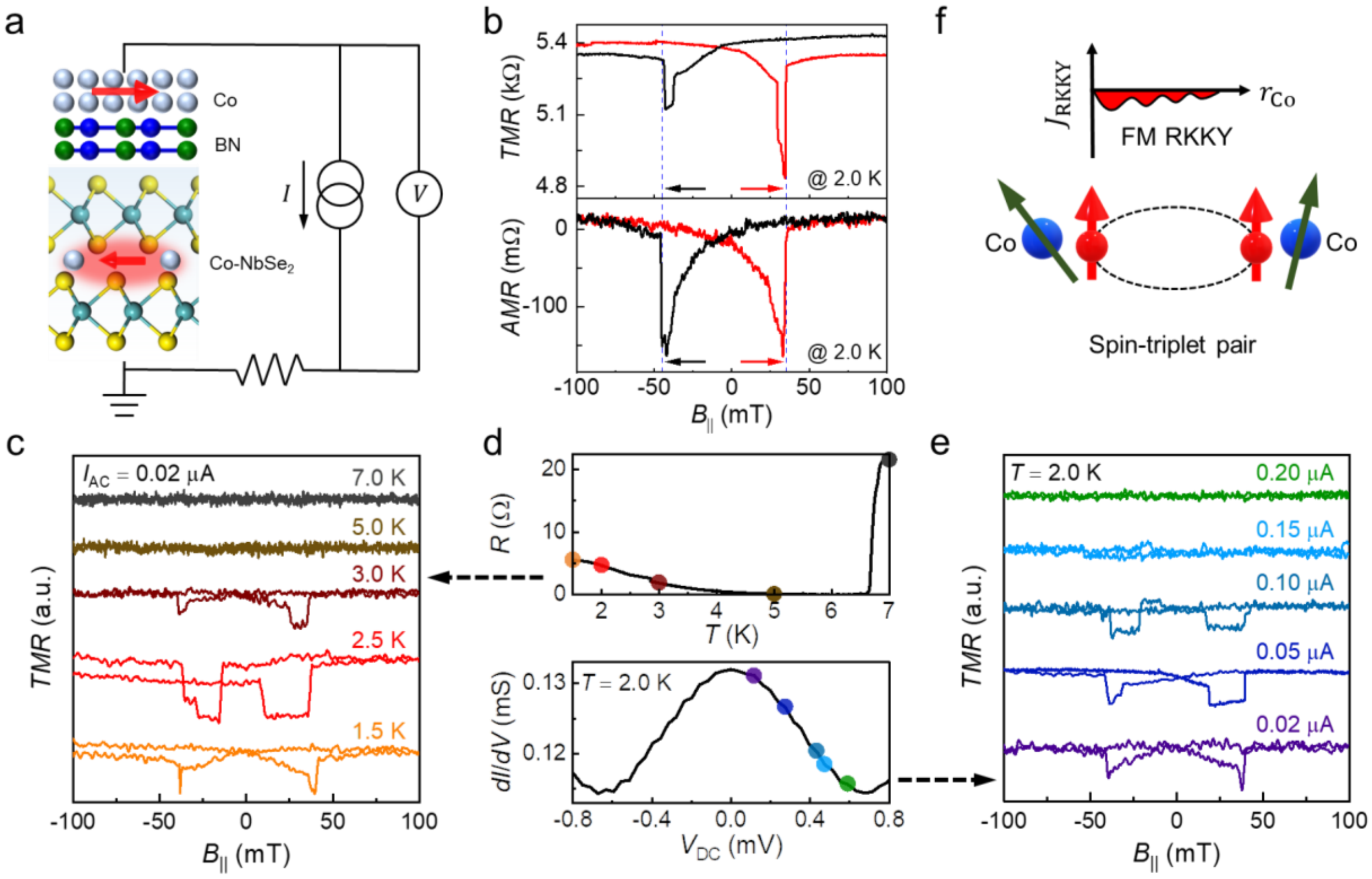

**Fig. 3 | Direct spin-dependent evidence for superconductivity-mediated ferromagnetism. a**, MTJ measurement schematics. Co and Co-$NbSe_2$ act as two ferromagnets that are separated by the thin BN tunnelling barrier. The TMR is measured as a function of $B_{||}$ along the easy axis of the Co contact. **b**, A comparison between the TMR of Co/BN/Co-$NbSe_2$ and the AMR of Co contact at $T = 2.0$ K. The TMR and AMR are measured with two separate AC currents at 0.1 μA and 10 μA, respectively. The switching in TMR is up to 500 Ω while the most prominent AMR effect is around 0.15 Ω. **c-e**, Temperature (**c**) and bias-current (**e**) dependences of the TMR signals through Co-$NbSe_2$ MTJ. Zero-field $R - T$ curve (top panel) and $dI/dV$ curve inside the superconducting gap (bottom panel) at $T =$ 2.0 K are shown in (**d**). The labelled coloured dots in the top (bottom) panel in (**d**) correspond to the five different temperatures (biases) in (**c**) and (**e**), respectively. Signature of ferromagnetism (the switching in TMR) only exists below $T_K$ and with the energy inside $\Delta_{SC}$. **f**, Schematics of a ferromagnetic RKKY (FM RKKY) interaction with spin-triplet pairs as the exchange carriers. A ferromagnetism is always preserved because of the exchange by spin-triplet pairs.

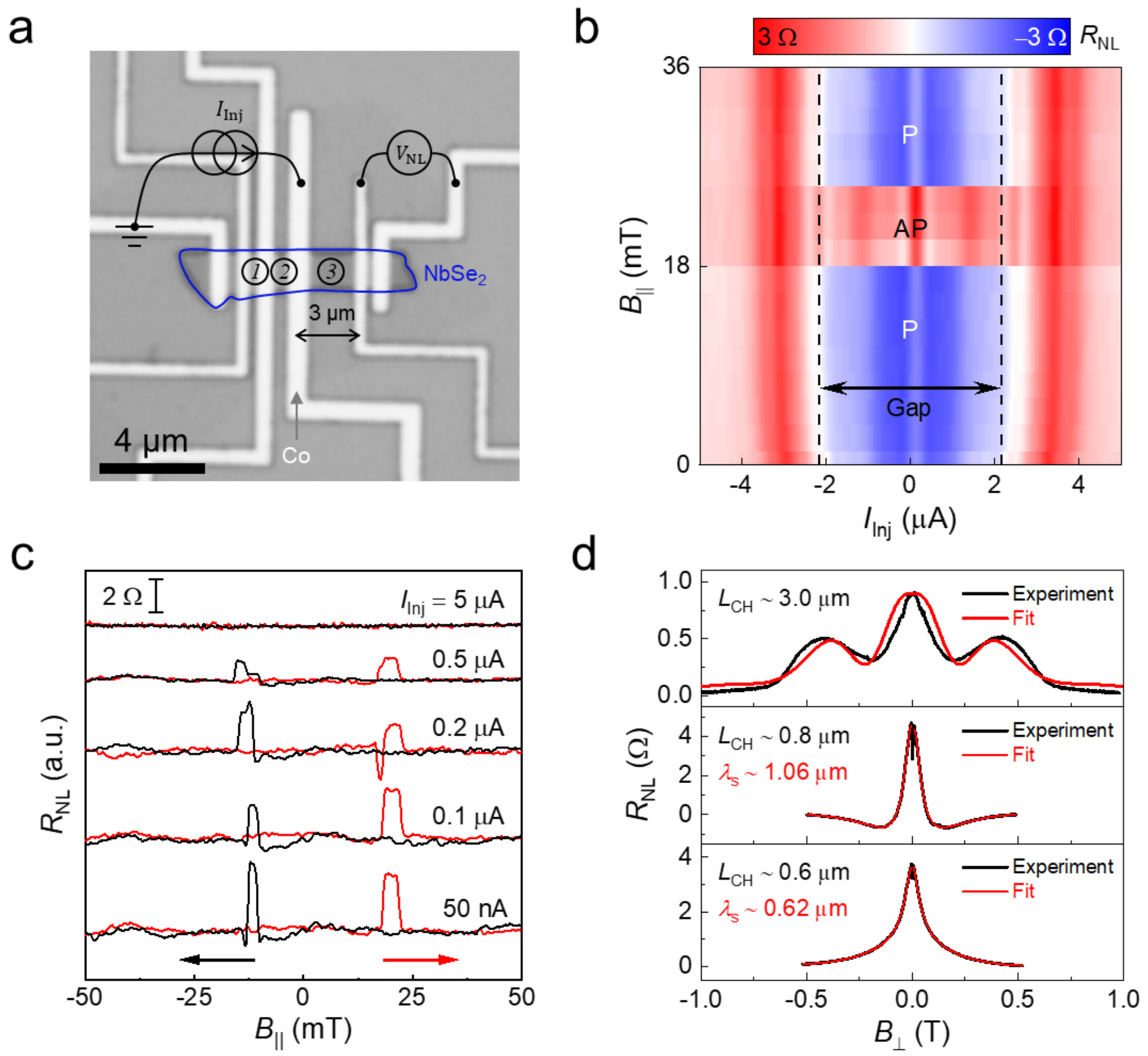

**Fig. 4 | Non-local spin supercurrent and Hanle precession. a**, Micrograph of the device for non-local measurement. The three different channels are labelled by #1 (~0.6 μm), #2 (~0.8 μm) and #3 (~3.0 μm), respectively. **b**, 2D mapping of $R_{NL}$ as a function of $B_{||}$ and $I_{Inj}$ at $T$ = 53 mK. A sign change of $R_{NL}$ occurs at finite $B_{||}$ near the coercive field of the Co detector and at finite $I_{Inj}$ for the potential inside the gap. A DC bias is swept from –5 μA to +5 μA with 0.05 μA in each step as the injection source and an AC bias equal to 0.05 μA is applied to resolve the non-local signal with a Lock-in amplifier. **c**, Examples of non-local switches with different $I_{Inj}$ at $T$ = 53 mK. The injection current tunes the magnitude of the switching but does not obviously affect the switching fields. **d**, $R_{NL}$ by sweeping $B_\perp$ at $T$ = 46 mK, with $L_{CH}$ = 0.6, 0.8 and 3.0 μm, respectively. The red lines are fitted by our modified Hanle precession model.

## Acknowledgements

B. Ö. acknowledges the support by the National Research Foundation, Prime Minister's Office, Singapore, under its NRF Investigatorship (Grant No. NRF-NRFI2018-8) and Medium-Sized Centre Programme. D-F-C. W. and S. A. would like to acknowledge the Singapore National Science Foundation Investigator Award (Grant No. NRF-NRFI06-2020-0003). F.H. and J.L. would like to acknowledge the support from National Natural Science Foundation of China (Grant No.11974156), the Science, Technology and Innovation Commission of Shenzhen Municipality (No. ZDSYS20190902092905285 and KQTD20190929173815000), and also the assistance of SUSTech Core Research Facilities, especially technical support from Cryo-EM Center and Pico-Centre that receives support from Presidential fund and Development and Reform Commission of Shenzhen Municipality. K.W. and T.T. acknowledge support from the Elemental Strategy Initiative conducted by the MEXT, Japan and the CREST (JPMJCR15F3), JST. We thank Alexander Hamilton, Feixiang Xiang, Manfred Sigrist and Zijin Lei for the fruitful discussions.

## Author Contributions

B. Ö. initiated, coordinated, and supervised the work. T. Q., D. F. and J. H. fabricated the devices and performed transport measurements. T. Q., B. Ö. and S. J. performed data analysis. X. C. assisted the data analysis. F. H. and J. L. performed the DF-TEM, HR-TEM and STEM characterizations. S. J., D-F-C. W. and S. A. provided theory work. K.W. and T.T. grew the BN crystals. T. Q., B. Ö., and S. J. co-wrote the manuscript.

## Competing Interests

The authors declare no competing interests.

## Methods

**Device design and fabrication.** The $NbSe_2$ flakes are mechanically exfoliated from bulk crystals (grown by HQ Graphene) by Scotch tapes onto a $Si/SiO_2$ substrate. The typical geometry of the $NbSe_2$ is about 10-μm long, and a 1~2-μm wide. The high-quality h-BN (grown by K. Watanabe and T. Taniguchi from National Institute for Materials Science, Japan) is exfoliated onto a photoresist-coated (PMMA 495-A4) Si wafer for the ease of hunting ultra-thin layers (2~4 layers) via optical contrast. The ultra-thin h-BN is then stacked onto the $NbSe_2$ by a semi-dry transfer method. Here, the ultra-thin BN layer plays three roles. First, it serves as an encapsulation layer that protects $NbSe_2$ from oxidation. Second, it is the diffusion barrier where atomic defeats are induced and enables the implantation of dilute ferromagnetism into $NbSe_2$. Third, at least one layer of BN remains relatively intact after the doping process, which plays an important role as the tunnelling barrier in the magnetic tunnel junction. The BN/$NbSe_2$ stack then undergoes a high-vacuum ($< 10^{-6}$ Torr) thermal annealing (210 ºC for 6 hrs) to remove the bubbles formed during the transfer. An e-beam lithography with acceleration voltage of 30 kV is employed to pattern the contacts, followed by an e-beam evaporation of Co/Ti films (35 nm/7.5 nm, deposition rate: 0.5 Å/s) under an ultra-high vacuum condition ($5\times10^{-8}$ Torr). We leverage the e-beam radiation during the pattern writing to induce short-range disorders in the ultra-thin BN and creates atomic defects in $NbSe_2$. And the evaporated Co atoms carry sufficient kinetic energy to diffuse through the disorders in BN and intercalate into $NbSe_2$. Au contacts are deposited separately via a thermal evaporator. The heterostructure goes through another high-vacuum thermal annealing to improve the quality of all contacts.

**Materials characterization.** The cross-sectional STEM specimens were prepared using a Cryo-focused Ion Beam in ultra-high vacuum ($< 10^{-6}$ mbar) in a liquid-nitrogen temperature environment. STEM imaging, EDS and EELS analysis of vertical heterojunctions of Co/BN/$NbSe_2$ were performed on an FEI Titan Themis with an X-FEG electron gun and a DCOR aberration corrector operating at 300 kV, at room temperature. The inner and outer collection angles for the STEM images ($\beta_1$ and $\beta_2$) were 48 and 200 mrad, respectively, with a convergence semi-angle of 25 mrad. The beam current was 100 pA for imaging and spectrum collection. In the elemental analysis, the Co concentration ($n_{\mathrm{Co}}$ in Fig. 1b) is defined as the ratio of detected Co atoms to the total atoms of all the detected elements. This ratio and its error bars were calculated using a quantitative analysis model of EDS and EELS spectrum with commercial codes that are standard and routinely used in this type of elemental assessment.

**Low-temperature measurement.** The electrical transport measurements were performed in a variable temperature insert (with a magnetic field up to 10 T and temperature down to 1.5 K for Device A) and a dilution refrigerator (with an in-plane magnetic field up to 7 T and out-of-plane magnetic field up to 2 T and temperature down to 40 mK for Devices B, C, D and E). A lock-in amplifier (with a frequency of 13 Hz) and a source measure unit (Keithley) were applied for the output and acquisition of the AC and DC signals, respectively. A four-probe measurement is used to analyse the $V-I$ and $R-T$ characteristics, a two-probe measurement is used to analyse the differential conductance and TMR signals, and a non-local measurement is used to characterize the spin transport. The in-plane magnetic field is applied along the easy-axis of the Co contacts. For the characterization of the AMR of Co contacts, a four-probe measurement is performed, where the magnetic field is parallel/antiparallel to the bias current through the easy axis of the Co contact.

**Superconducting gap symmetries and YSR states.** To analyse the temperature and magnetic field dependences of superconducting gap and YSR states, we fit our differential conductance data with a modified Bardeen–Cooper–Schrieffer (BCS) theory[32],[44],

$$\frac{dI}{dV}(V) = \frac{\sigma_N}{2\pi}\int_{-\infty}^{\infty} dE \left(\frac{df(\epsilon)}{d\epsilon}|_{\epsilon=E-eV}\right)\int_0^{2\pi} d\theta\, Re\left[\frac{E - i\Gamma}{\sqrt{(E - i\Gamma)^2 - \Delta_0 \cos^2(l\theta)}}\right] + \sigma_{ZBP}\, exp\left(-\frac{V^2}{2\sigma^2}\right), \quad (1)$$

where $\sigma_N$ is the background normal state differential conductance, $\Gamma$ is the quasiparticle broadening and $\Delta_0 \cos(l\theta)$ is the superconducting gap size with the angular-dependent factor $l$ ($l = 0, 1, 2, 3$ for $s$-, $p$-, $d$-, and $f$-wave superconducting order parameters, respectively). We only consider $l = 0$ and 1 for singlet and triplet fittings. The second term in Equation (1) is added to contain the YSR states.

**Superconducting coherence length.** To analyse the suppression of $T_C$ by magnetic field, we introduce the pair-breaking equation[45] under an external magnetic field. Since the thickness is much thinner than the parallel penetration depth $\lambda_{||} \approx 69$ nm[46], we apply the pair-breaking equation in thin film,

$$\ln\frac{T_C}{T_{C0}} = \Psi\left(\frac{1}{2}\right) - \Psi\left(\frac{1}{2} + \frac{\alpha}{2\pi k_B T_C}\right), \quad (2)$$

where $\alpha = De^2 B_{\parallel}^2 d^2/6\hbar$ for parallel field and $\alpha = DeB_{\perp}$ for perpendicular field. The superconducting coherence length $\xi_{GL}^0 = \xi_{GL}(T = 0, B = 0)$ can be extracted by the electronic diffusion constant $D = 16k_B T_{c0}\xi_{GL}^0/h$[47]. The fitting results by Equation (2) are shown in Fig. 1d, and the extracted values of $\xi_{GL}^0$ are shown in Extended Data 10.

**Hanle precession and non-local resistance.** For the Hanle precession measurement, the injector and detector are first polarized in parallel or antiparallel directions to each other by an in-plane magnetic field. Then the in-plane magnetic field is cancelled, and an out-of-plane field to superconducting flake is performed. For $L_{CH} = 0.6$ and 0.8 µm, we follow the standard Hanle model[48] below,

$$R_{NL} = R_0 \int_0^{\infty} dt \frac{1}{\sqrt{4\pi Dt}} e^{-\frac{L_{CH}^2}{4Dt}} e^{-\frac{t}{\tau}} \cos(\omega_L t), \quad (3)$$

where $R_0$ is a scale factor for the resistance, $D$ is the spin diffusion constant, $L_{CH}$ is the centre-to-centre distance between the injector and detector, $\tau$ is the spin relaxation time, $\omega_L = g\mu_B B_{\perp}/\hbar$ is the Larmor frequency with Landé $g$-factor ($g = 2$), $\mu_B$ is the Bohr magneton and $\hbar$ is the Plank constant. The spin relaxation lengths are computed by $\lambda_s = \sqrt{D\tau} \sim 0.62$ and 1.06 µm for the two channels, respectively.

However, for $L_{CH} = 3.0$ µm, the standard Hanle model cannot describe the second peak (without a sign change) at $B_{\perp} \approx \pm 0.4$ T, which suggests that Meissner effect should be taken into account (given $NbSe_2$ is a Type-II superconductor). Therefore, we introduce a term incorporating the vortices in the superconducting channel, and the modified Hanle model is in the form below,

$$R_{NL} = R_0 \int_0^{\infty} dt \frac{1}{\sqrt{4\pi Dt}} e^{-\frac{L_{CH}^2}{4Dt}} e^{-\frac{t}{\tau}} [1 - a_v + a_v \cos(\omega_L t)], \quad (4)$$

where $a_v = 1 - \exp(-B/B_v)$ is a parameter accounting for the proportion of the transverse area with vortices responsible for the screening effect of magnetic field, and $B_v$ is a measure of the magnetic field responsible for the area density of vortices.

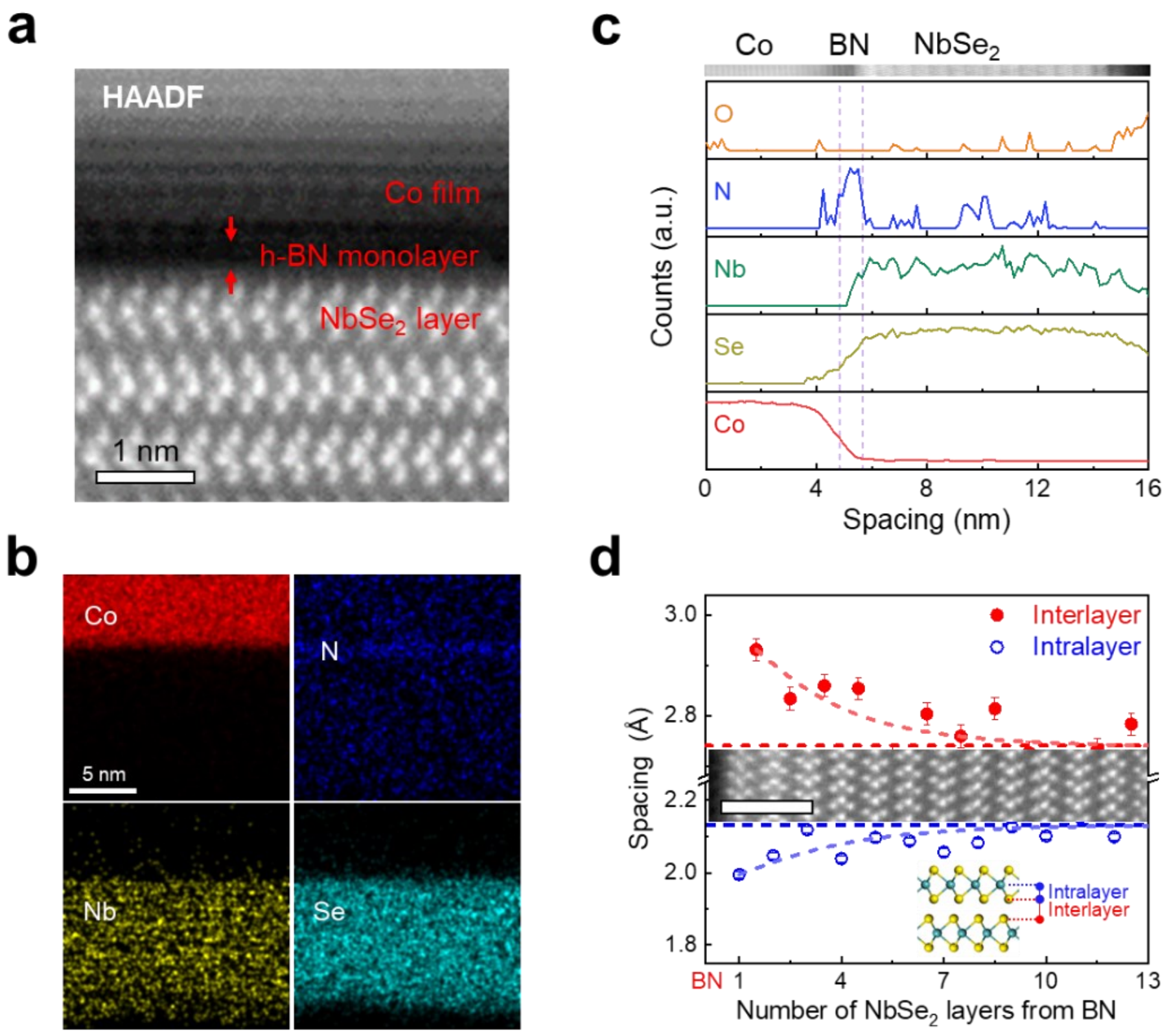

**Extended Data 1 | TEM characterizations of Co intercalation.** **a,** Cross-section STEM image of Co-intercalated $NbSe_2$ heterojunction. The BN region is determined by electron energy loss spectra (EELS), where a crystalline monolayer or bilayer BN always remains after the deposition of Co. **b,** Energy dispersive spectrum (EDS) maps for the elements Co, N, Nb, Se. Scale bar: 5 nm. **c,** Line cuts of the elemental abundance perpendicular to the Co/BN/Co-$NbSe_2$ interface, extracted from EELS. **d,** Evolution of the VdW interlayer gap (red) and intralayer gap (blue), extracted from STEM cross-sections (inset), as a function of distance to the BN/$NbSe_2$ interface (in number of $NbSe_2$ monolayers). The horizontal red and blue dashed lines respectively refer to the interlayer and intralayer gaps in the pristine bulk $NbSe_2$. Scale bar: 2 nm. Inset: Schematic definition about the intralayer and interlayer.

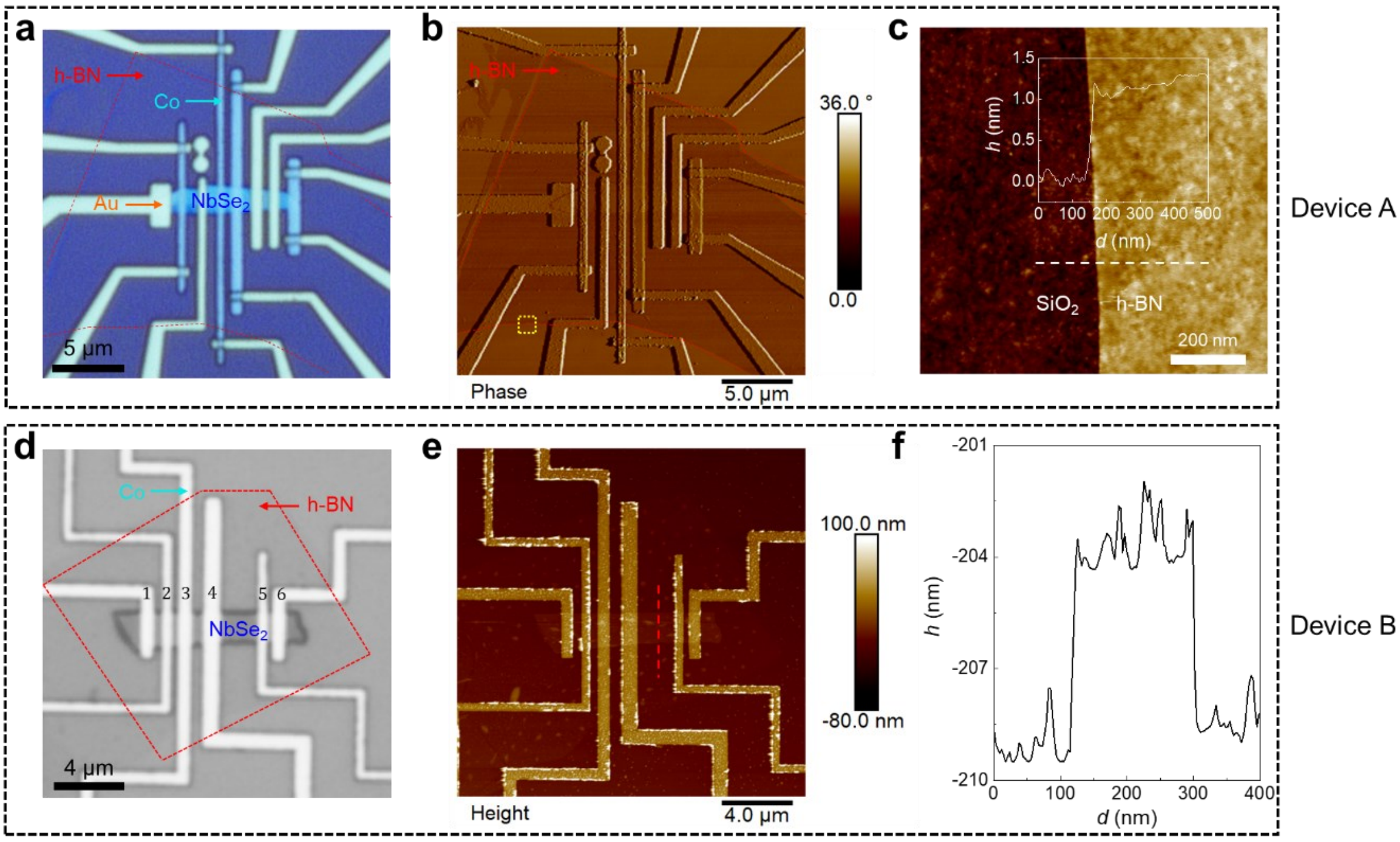

**Extended Data 2 | Device morphology. a,** Optical micrograph of Device A. The Co and Au contacts are deposited onto the BN-encapsulated channel ($NbSe_2$) in an alternate pattern, where both the Co-doped and -undoped SC regions can be characterized. The easy axes of the Co contacts are aligned in cross with the channel for the ease of spin injection into the channel. **b,** The AFM image of (**a**) in phase contrast. **c**, Thickness profile of the BN flake (the probed region is denoted by the yellow dashed box in **b**). **d,** Optical micrograph of Device B. **d**, Optical micrograph of Device B. An optical filter (wavelength ~ 560 nm) is applied to increase the outline contrast for BN and $NbSe_2$. For Hanle precession measurements, the three characterized channels (from the bottom panel to the top panel in Fig. 4d) are between contacts #2 and #3, #3 and #4, and #4 and #5, respectively. **e**, The AFM image of (**d**). **f**, Thickness profile of the $NbSe_2$ flake (the probed region is denoted by the red dashed line in **e**).

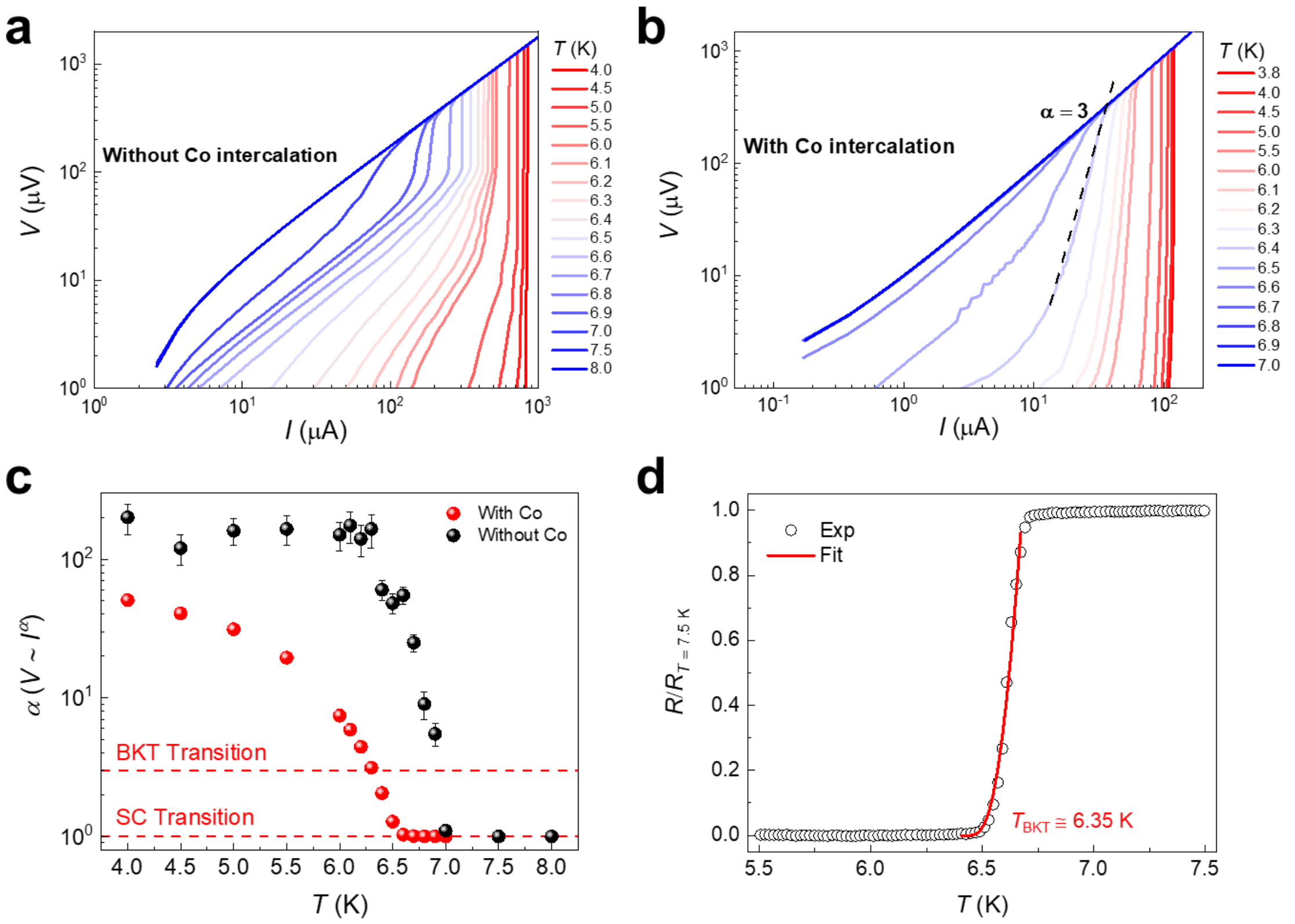

**Extended Data 3 | BKT transition in Co-intercalated $NbSe_2$ and the fitting models. a, b,** $V-I$ curves for non-Co-intercalated (**a**) (Device D) and Co-intercalated (**b**) $NbSe_2$ (Device A) with temperatures ranging from base to just above $T_C$. **c**, Temperature dependence of the exponent $\alpha$ deduced from the power-law, $V \sim I^{\alpha}$. As indicated by the red dashed line, $\alpha$ approaches 3 at $T = 6.3$ K for the Co-intercalated sample. For the non-Co-intercalated sample, $\alpha$ shows an abrupt jump from around 1 to around 6 from $T_C$ to just below $T_C$. **d**, Temperature dependence of the resistance at zero field. With the condition $T_{BKT} < T < T_C$, $R(T)$ follows the relation in the form of $R = R_0 \cdot e^{-a\hat{T}^{-1/2}}$, where $R_0$ and $a$ are parameters reflecting the properties of the specific superconductor and $\hat{T} = T/T_{BKT} - 1$ is the reduced temperature. The red line is a fit to BKT transition, yielding $T_{BKT} \approx 6.35$ K.

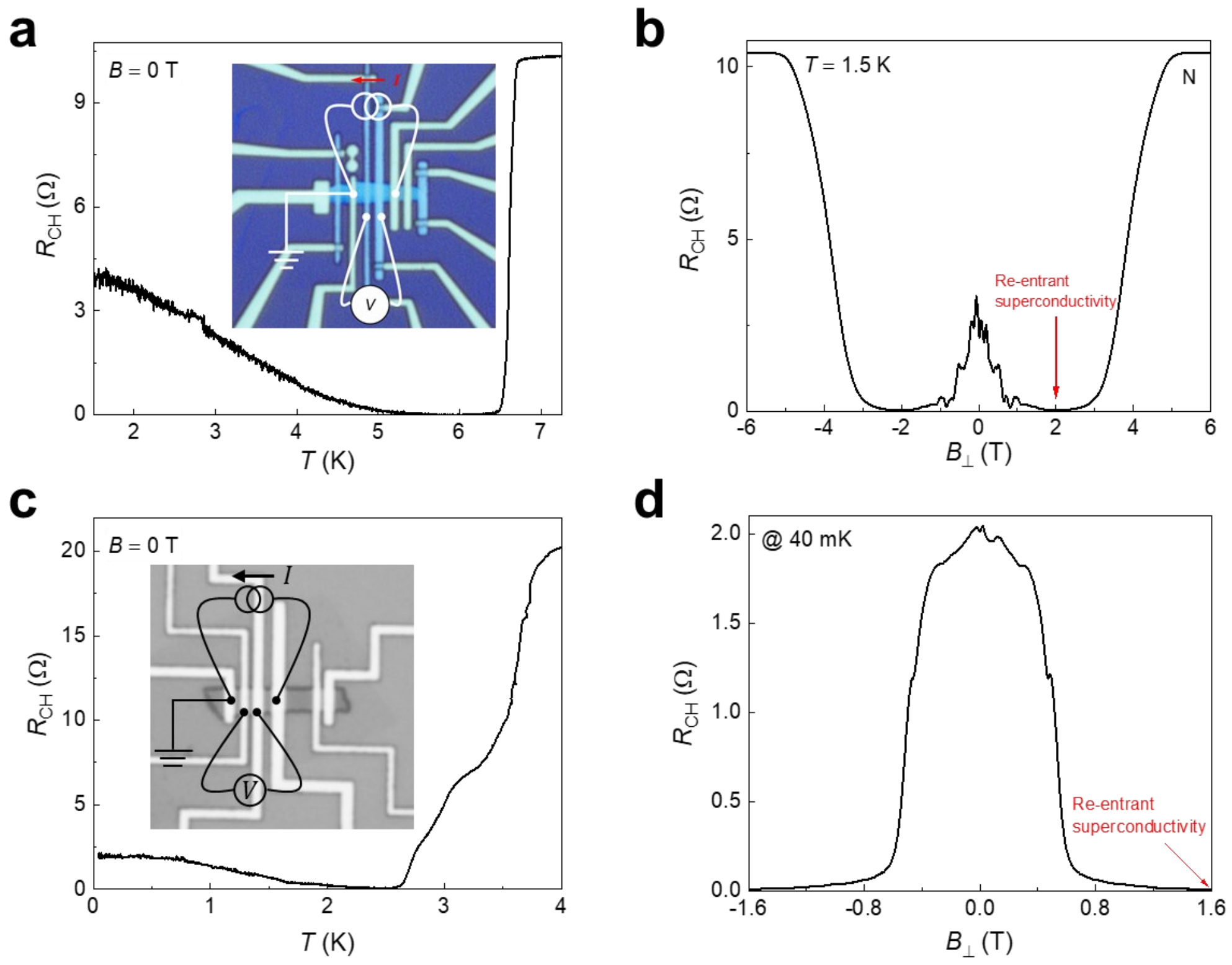

**Extended Data 4 | Re-entrant superconductivity under out-of-plane field. a,** $R-T$ curve at zero magnetic field for Device A. Inset: measurement set-up. **b,** $R-B_{\perp}$ curve at $T = 1.5$ K for Device A. A re-entrant superconductivity occurs at around 2 T and the system enters the normal state at around 5 T. **c,** $R-T$ curve at zero magnetic field for Device B. Inset: measurement set-up. **d,** $R-B_{\perp}$ curve at $T = 80$ mK for Device A. A re-entrant superconductivity occurs at around 1.6 T.

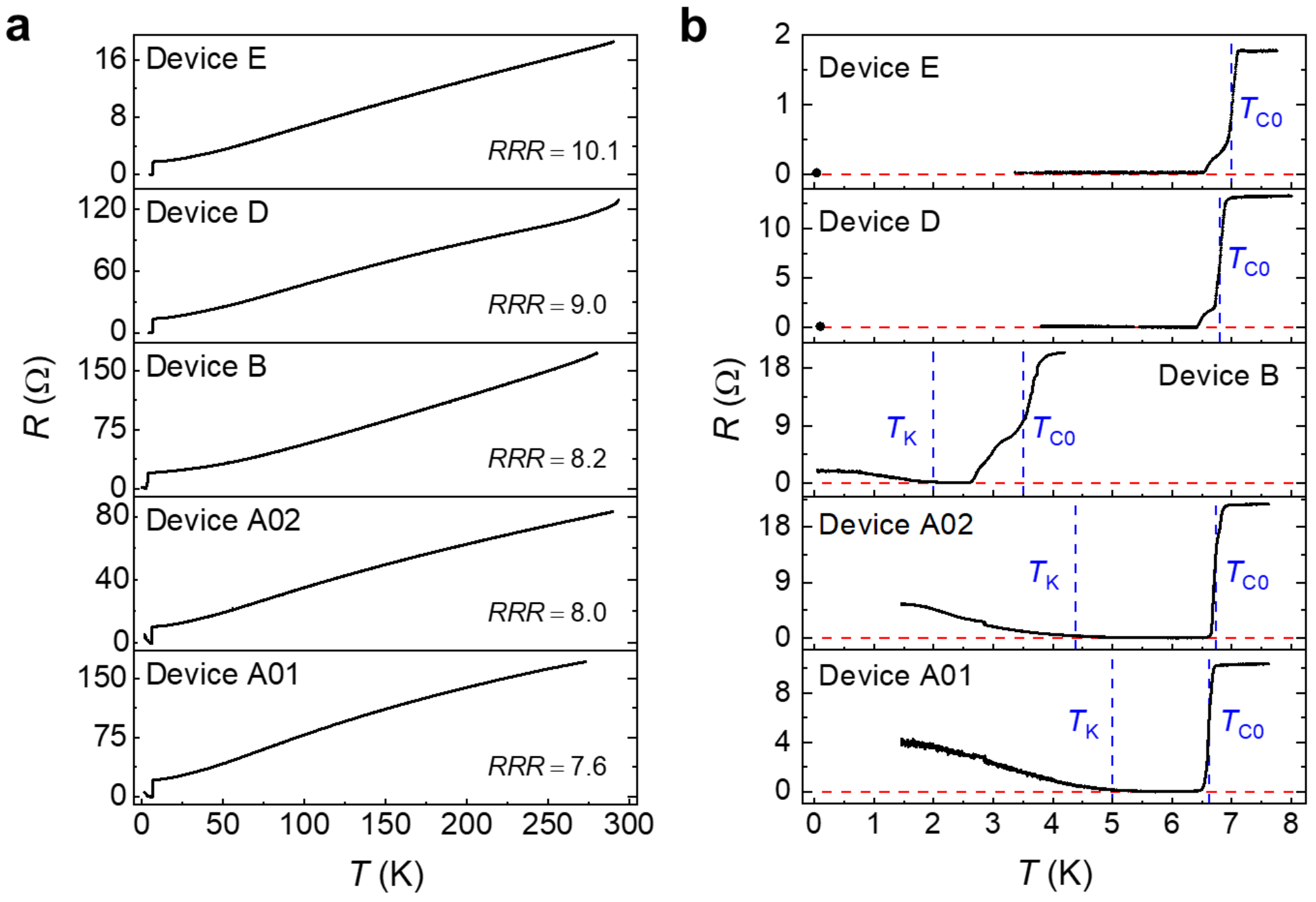

**Extended Data 5 | Doping control of the $RRR$, $T_K$ and $T_C$. a,** $R-T$ curves of several devices with different Co doping levels from 273 K to base temperature. The $RRR$ is defined as $RRR = R_{273K}/R_{T_C^+}$, where $R_{T_C^+}$ refers to the resistance just above $T_C$. **b,** $R-T$ curves of the corresponding devices in (**a**) from $T_C^+$ to base temperature. The red dashed lines are references for zero resistance. For all the four devices, a constant AC current (1 μA) is applied through the channel for the measurement. We note that for the ferromagnetic phase to emerge, a critical Co intercalation is needed. For the two devices (A and B) showing the resistance anomaly and spin switching have similar Co doping levels estimated by the *RRR* (with the value around 7.6 ~ 8.2, which corresponds to average Co concentration around 3~5% by STEM). For the devices without sufficient Co doping, we do not observe any signature of superconductivity-mediated ferromagnetism, and the *RRR* is no less than 9.

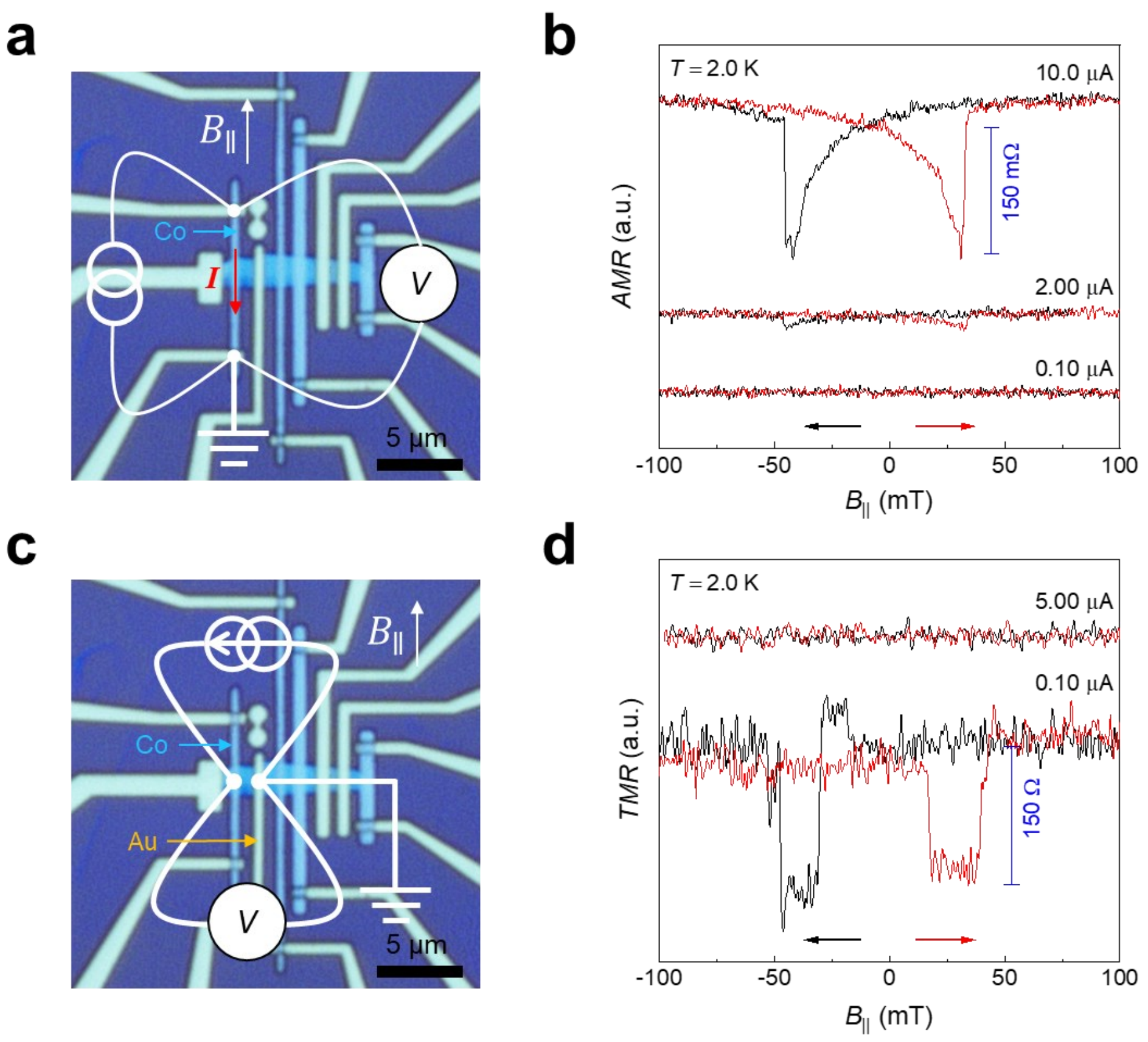

**Extended Data 6 | Comparison between the AMR of Co contact and the TMR of in Co/BN/Co-NbSe$_2$ MTJ. a,** Measurement set-up for Co AMR. A constant current is passed along the easy axis of the Co stripe where the magnetic field is swept. **b,** Bias-current dependence of the AMR signals of the Co contact. The AMR is only prominent if the bias current is large (> 2 μA) and its magnitude is in mΩ scale. **c,** Measurement set-up for Co/BN/Co-NbSe$_2$ TMR. A constant current is through the Co/BN/Co-NbSe$_2$ and a non-ferromagnet (Au) is used as ground. **d,** Bias-current dependence of the TMR signals of the MTJ. The TMR is only prominent if the bias inside $\Delta_{SC}$ and the magnitude is in 100-Ω scale.

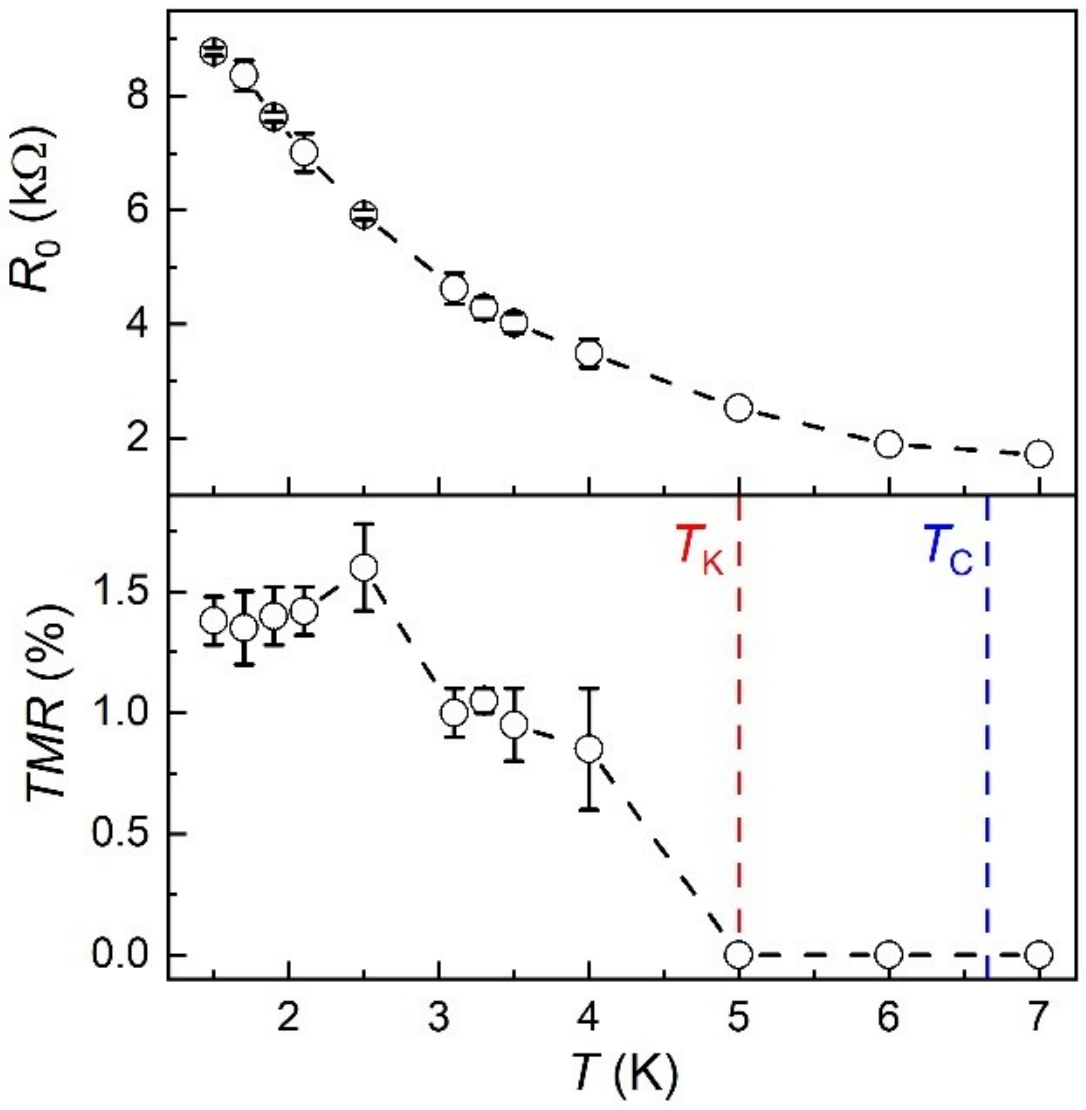

**Extended Data 7 | Temperature dependence of the two-probe MTJ resistance at zero field ($R_0$) and the switching magnitude in TMR.** The $R_0$ increases with cooling down below $T_C$ due to the enhanced $\Delta_{SC}$. The TMR (%) only emerges below $T_K$ and saturates at around 1.5% below 2.5 K. $I_{AC} =$ 0.02 μA.

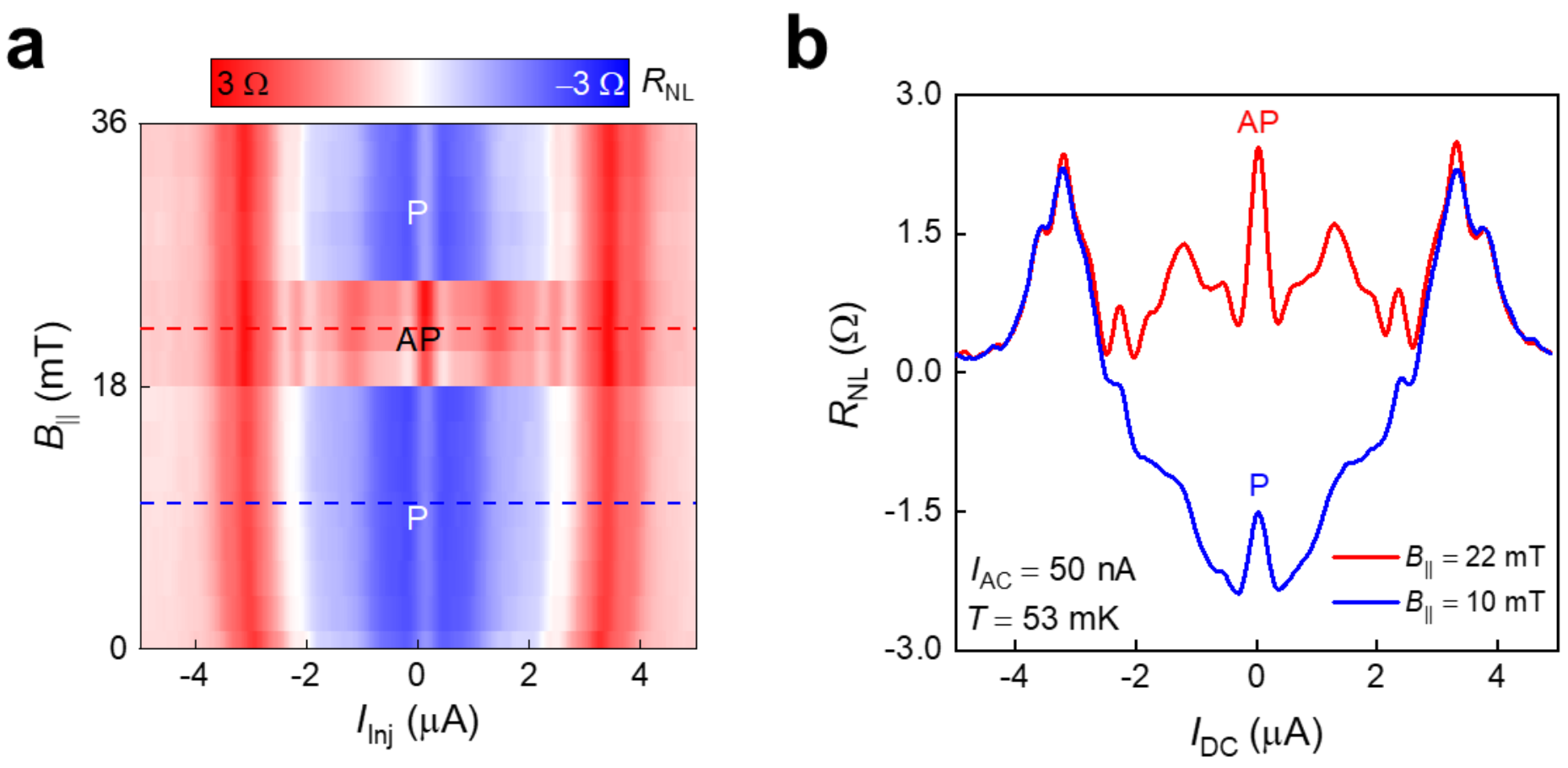

**Extended Data 8 | Hanle precession signals in the superconducting spin valve. a,** 2D mapping of $R_{\mathrm{NL}}$ as a function of $B_{||}$ and $I_{\mathrm{Inj}}$. **b,** $R_{\mathrm{NL}}$ as a function of $I_{\mathrm{Inj}}$ under two discrete magnetic fields highlighted by the blue dashed line (parallel regime) and red dashed line (anti-parallel regime) in (**a**).

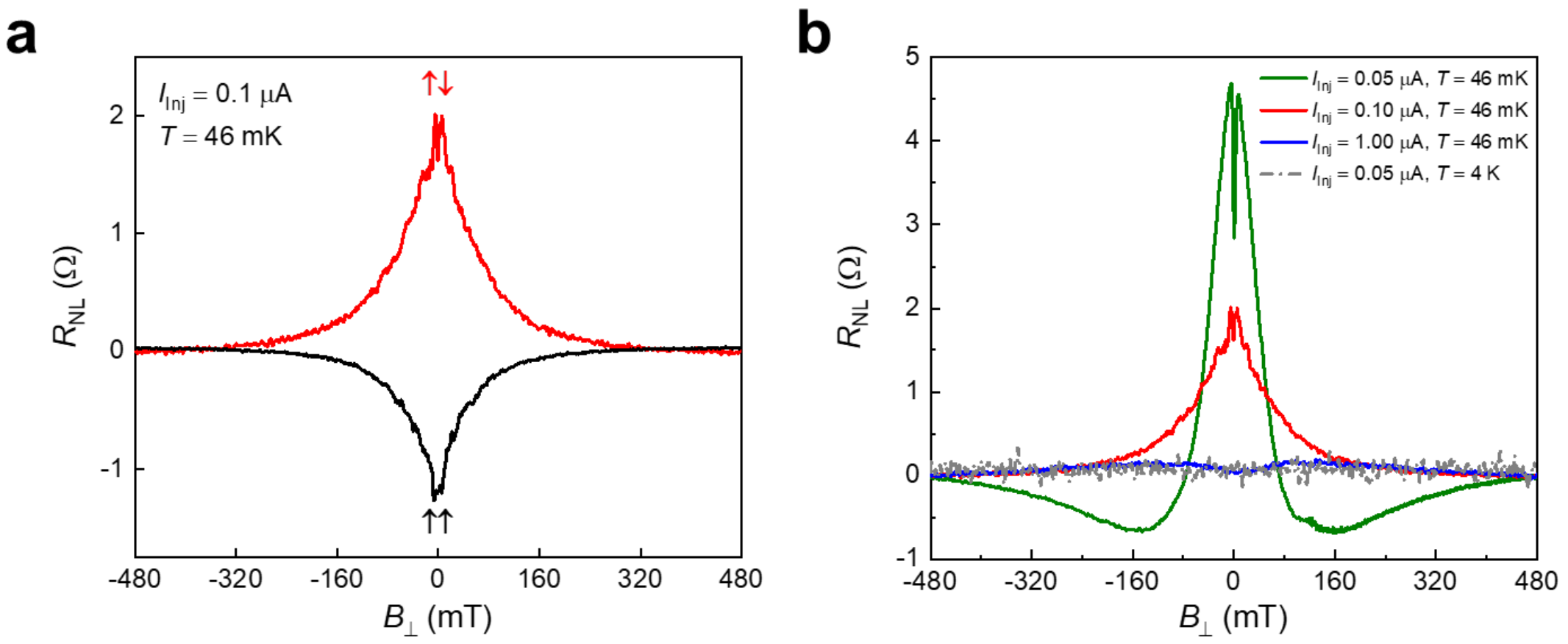

**Extended Data 9 | Hanle precession measurements in the superconducting spin valve. a,** $R_{\mathrm{NL}}$ as a function of the out-of-plane magnetic field with $I_{\mathrm{Inj}} = 0.1$ μA. The opposite signs of the two peaks in the precession curves correspond to the parallel and antiparallel configurations between the injector and detector. **b,** Injection bias current ($I_{\mathrm{Inj}}$) dependence of the Hanle precession signals at 46 mK. The grey dashed line is taken at $I_{\mathrm{Inj}} = 0.05$ μA and $T = 4.0$ K ($> T_{\mathrm{C0}}$).

| Device ID | $NbSe_2$ thickness (nm) | BN layers | Contacts | $T_{C0}$ (K) | $\xi_{GL}^{0}$ (nm) |
|---|---|---|---|---|---|
| A01 | ~12 | 3 | Au−Co−Co−Au | 6.60 | 5.7±1.2 |
| A02 | ~12 | 3 | Au−Co−Au−Au | 6.65 | 5.6±1.2 |
| B | < 4 | 3 | Co−Co−Co−Co | 3.50 | NA |
| C | ~10 | ≤2 | Co−Co−Co−Co | 4.70 | 7.1±1.4 |
| D | ~10 | ~5 | Au−Co−Co−Au | 6.80 | NA |
| E | ~10 | 3 | Au−Au−Au−Au | 6.95 | 3.7±0.7 |

**Extended Data 10 | Basic parameters about our devices.**